# Distinct band reconstructions in kagome superconductor $CsV_3Sb_5$

Yang Luo[1,#], Shuting Peng[1,#], Samuel M. L. Teicher[2,#], Linwei Huai[1,#], Yong Hu[1,3], Brenden R. Ortiz[2], Zhiyuan Wei[1], Jianchang Shen[1], Zhipeng Ou[1], Bingqian Wang[1], Yu Miao[1], Mingyao Guo[1], M. Shi[3], Stephen D. Wilson[2] and J.-F. He[1,*]

[1]*Hefei National Laboratory for Physical Sciences at the Microscale, Department of Physics and CAS Key Laboratory of Strongly-coupled Quantum Matter Physics, University of Science and Technology of China, Hefei, Anhui 230026, China*

[2]*Materials Department and California Nanosystems Institute, University of California Santa Barbara, Santa Barbara, California 93106, USA*

[3]*Swiss Light Source, Paul Scherrer Institute, CH-5232 Villigen PSI, Switzerland*

[#]These authors contributed equally to this work.

*To whom correspondence should be addressed: jfhe@ustc.edu.cn

**The new two-dimensional (2D) kagome superconductor $CsV_3Sb_5$ has attracted much recent attention due to the coexistence of superconductivity, charge order, topology and kagome physics [1-33]. A key issue in this field is to unveil the unique reconstructed electronic structure, which successfully accommodates different orders and interactions to form a fertile ground for emergent phenomena. Here, we report angle-resolved photoemission spectroscopy (ARPES) evidence for two distinct band reconstructions in $CsV_3Sb_5$. The first one is characterized by the appearance of new electron energy band at low temperature. The new band is theoretically reproduced when the three dimensionality of the charge order [21-23] is considered for a band-folding along the out-of-plane direction. The second reconstruction is identified as a surface induced orbital-selective shift of the electron energy band. Our results provide the first evidence for the three dimensionality of the charge order in single-particle spectral function, highlighting the importance of long-range out-of-plane electronic correlations in this layered kagome superconductor. They also point to the feasibility of orbital-selective control of the band structure via surface modification, which would open a new avenue for manipulating exotic phenomena in this system, including superconductivity.**

CsV$_3$Sb$_5$ is a member of the new class of AV$_3$Sb$_5$ (A= K, Rb, Cs) kagome superconductors [1-36] with a layered crystal structure (Fig. 1a). The V sublattice forms a perfect kagome net (Fig. 1b), which is interwoven with a hexagonal net of Sb atoms in the same plane (Fig. 1a). This V$_3$Sb layer is then bounded above and below by Sb honeycomb lattices and Cs hexagonal lattices (Fig. 1a). The corresponding three dimensional (3D) Brillouin zone (BZ) is shown in Fig. 1c. Due to the layered crystal structure, a projected 2D BZ is often used for simplicity (see Fig. 1c,d; the high symmetry points are marked as $\bar{\Gamma}$, $\bar{K}$ and $\bar{M}$). While topological states and electron correlation effects are naturally expected from the kagome net in CsV$_3$Sb$_5$, the existence of a charge density wave (CDW) order below T$_{CDW}$=94K and a superconducting ground state (T$_c$=2.5K) has further increased the richness of this system at low temperature [2-33].

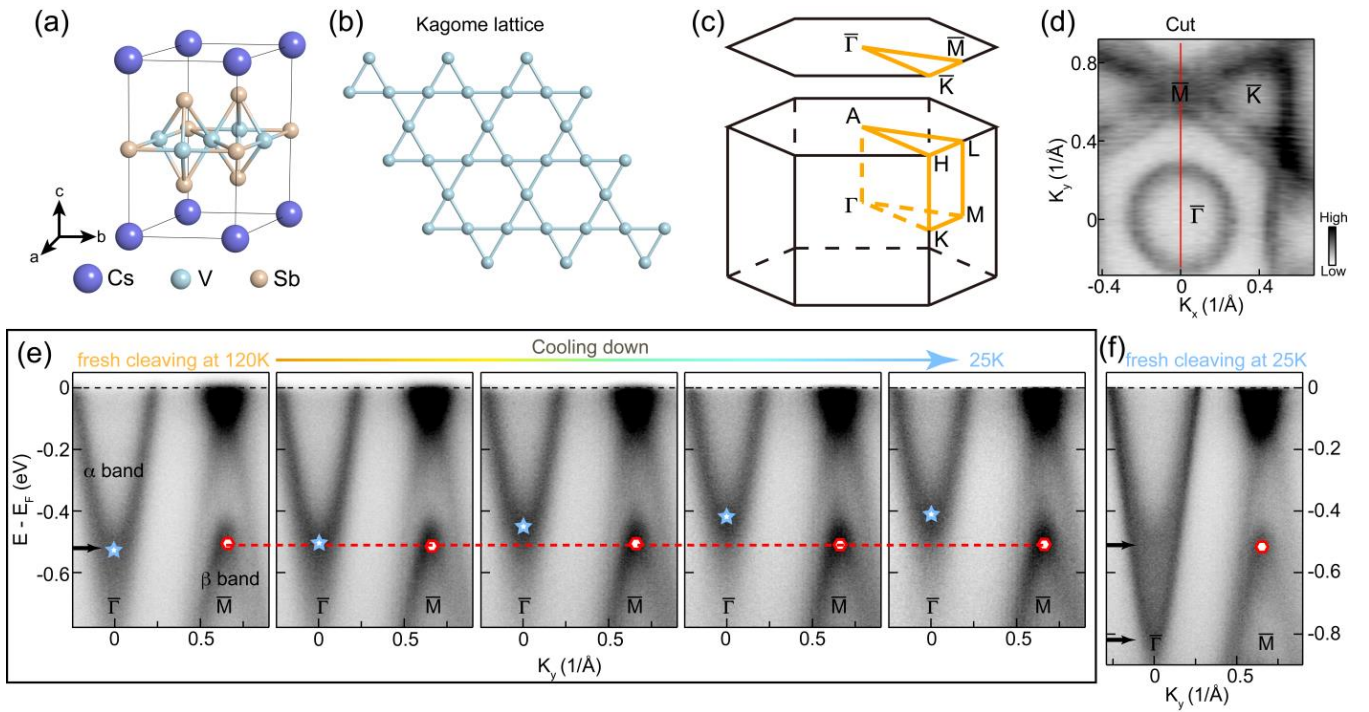

**Fig. 1 Distinct band reconstructions in CsV$_3$Sb$_5$.** *(a) Crystal structure of CsV$_3$Sb$_5$. (b) The kagome lattice formed by vanadium atoms. (c) Schematic of the 3D BZ and the projected 2D BZ. (d) Fermi surface mapping shown in the projected 2D BZ. (e) Band structure evolution during the cooling process. The location of the $\bar{\Gamma}-\bar{M}$ momentum cut is marked by the red line in (d). (f) Band structure along $\bar{\Gamma}-\bar{M}$ direction measured at 25K on a fresh sample surface. The blue stars (red hexagons) mark the bottom (top) of the α (β) band. The red dashed line is a guide for the eye.*

In order to track the evolution of the electronic structure, the sample was cleaved at a temperature above the CDW transition (T>$T_{CDW}$) and slowly cooled down towards a low temperature (T=25K). As shown in Fig. 1e, the electron-like band around $\bar{\Gamma}$ moves up towards the Fermi energy ($E_F$) (marked by blue stars), while the hole-like band around $\bar{M}$ remains at almost the same energy (marked by red hexagons). For simplicity, we label these two bands as α band and β band, respectively (Fig. 1e). However, the above band evolution is distinct from the electronic structure measured on a sample directly cleaved at the low temperature (25K). As shown in Fig.1f, the upward energy shift of the α band around $\bar{\Gamma}$ is absent. Instead, a clear electron-like band bottom is observed at a much deeper binding energy (around -0.8eV at the $\bar{\Gamma}$ point). The different band structures probed at the same temperature (25K) indicate the existence of two distinct band reconstructions in this material system.

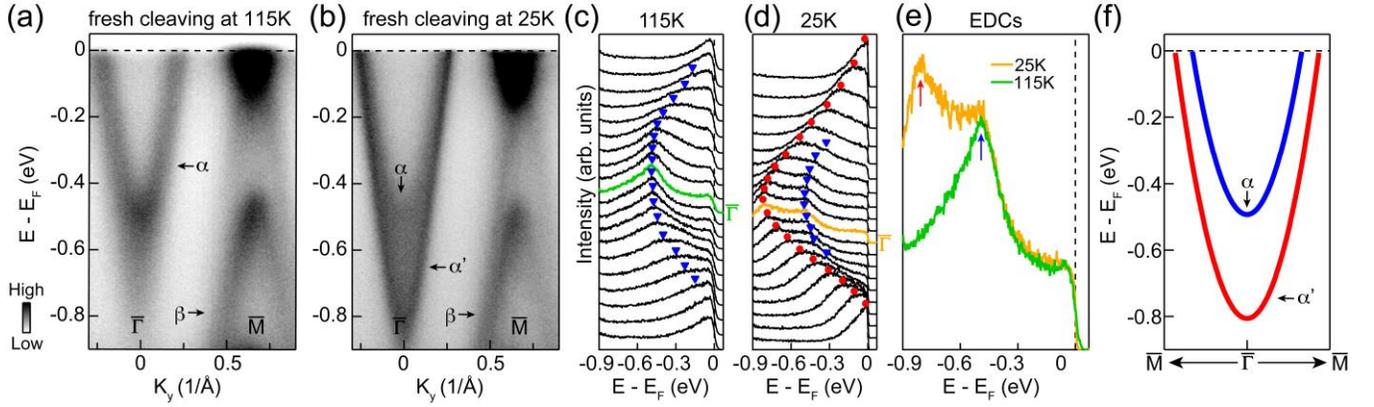

*Fig. 2 Comparison of the band structure above and below the CDW transition temperature on fresh sample surfaces. (a) Band structure along the $\bar{\Gamma}-\bar{M}$ direction measured at 115 K on a fresh sample surface cleaved at the same temperature. (b) Same as (a), but at 25K. (c-d) EDCs near $\bar{\Gamma}$ for the data measured at 115K (c) and 25K (d), respectively. Blue triangles label the α band, and red circles label the new α' band. The EDC at $\bar{\Gamma}$ is shown in green (yellow) for the data at 115K (25K), and plotted in an expanded scale in (e). The blue arrow in (e) marks the EDC peak from the α band, and the red arrow marks the peak from the new α' band at 25K. (f) Schematic of the α and α' bands. The blue curve represents the temperature independent α band, and the red curve represents the new α' band observed at 25K.*

To avoid possible complications during the cooling process, we first compare the band structure of two CsV$_3$Sb$_5$ samples cleaved at 115K (T>T$_{CDW}$) and 25K (T<T$_{CDW}$), respectively. The electron energy bands in the normal state (T>T$_{CDW}$) are well captured by the corresponding DFT calculations [2] (Fig. 4a). However, more electronic features seem to appear at low temperature with a prominent band reconstruction around $\bar{\Gamma}$ (Fig. 2b). A careful examination of the energy distribution curves (EDCs) reveals a new electron-like band (marked by the red circles in Fig. 2d) in addition to the original α band (marked by the blue triangles in Fig. 2c,d). This is quantitatively confirmed by the temperature dependence of the EDC at $\bar{\Gamma}$. As shown in Fig. 2e, the EDC peak from the α band (marked by the blue arrow) shows little change with temperature while an additional peak from the new band only appears at low temperature (around -0.8eV; marked by the red arrow; also see Supplemental Materials Fig. S1). For simplicity, we label the new electron-like band observed at low temperature as α' band, hereafter. We also summarize the original α band (blue) and the low temperature α' band (red) in Fig. 2f.

After establishing the temperature induced band reconstruction, we next investigate the other type of band evolution observed during the cooling process. To isolate the tuning parameters, we fix the temperature at 115K and measure the band structure as a function of time (Fig. 3). We track the energy position of α (β) band bottom (top) by the EDC peak at $\bar{\Gamma}$ ($\bar{M}$), as illustrated in Fig. 3a,b (d,e). The systematic time evolution of the EDC peak is shown in Fig. 3c (f). It is clear that the α band moves up towards the E$_F$, while the β band shows little change as a function of time. The overall band evolution reproduces that observed during the cooling process (Fig. 1e). Therefore, this type of band reconstruction is associated with time but not temperature.

We next discuss the origins of these two distinct band reconstructions in this system. First, we consider the band reconstruction as a function of temperature. The coexistence of both α and α' bands at low temperature seems to suggest a band splitting, which has been reported in magnetic materials [37]. However, CsV$_3$Sb$_5$ does not exhibit any resolvable magnetic order [2]. Moreover, a band splitting would change the energy position of the original α band, which is inconsistent with the experiment. Electron-phonon coupling can also give rise to a replica band at a deeper energy [38]. But the energy separation between α band and α' band is greater than 300meV, which is much higher than the phonon energies in this material system. A surface reconstruction has been reported to account for additional shadow bands in layered-materials [39]. But such a surface reconstruction has not been observed in $CsV_3Sb_5$ yet. It also remains unclear how surface reconstruction can give rise to the α' band at a deeper energy. On the other hand, we find that the temperature induced band reconstruction can be well reproduced by considering an out-of-plane band folding induced by the 3D charge order in $CsV_3Sb_5$ [17,21-24]. While the band structure of $CsV_3Sb_5$ is primarily quasi-2D, a significant $K_z$ difference can be identified between Γ and A in the 3D BZ (Fig. 4a), where the electron-like band around A locates at a much deeper energy. As shown in Fig. 4c, when the unit cell along the out-of-plane direction is doubled by the 3D charge order, the ALH plane in the BZ will be folded to the ΓMK plane. As such, the deep electron-like band along L-A-L (Fig. 4a) will be folded to the M-Γ-M direction (Fig. 4b), giving rise to the α' band observed in the experiment. We note that the DFT calculation also shows a hole-like band close to the bottom of the α' band. However, this band disappears when in-plane component of the charge order [19] or electron correlation [40] is considered in the calculations. In this sense, our results reveal the direct coupling of the real space 3D charge order [21,22] to the single particle spectral function of the system. We note that electronic states dressed by the long-range out-of-plane correlation of charge order in $CsV_3Sb_5$ may have important implications. For instance, high-temperature superconductivity in cuprates shows 3D superconducting coherence with primarily 2D electronic band structures [41], and the experimental identification of 3D charge order revealed an intimate link between superconductivity and charge order beyond simple competition [42]. In the current case, 3D superconductivity is realized in the quasi-2D kagome metal [2-11,33], where 3D charge order also appears and dresses the quasi-2D electronic structure. It would be interesting to explore the nature of the out-of-plane interaction that stabilizes the 3D ordering tendencies in $CsV_3Sb_5$ and related quasi-2D kagome superconductors.

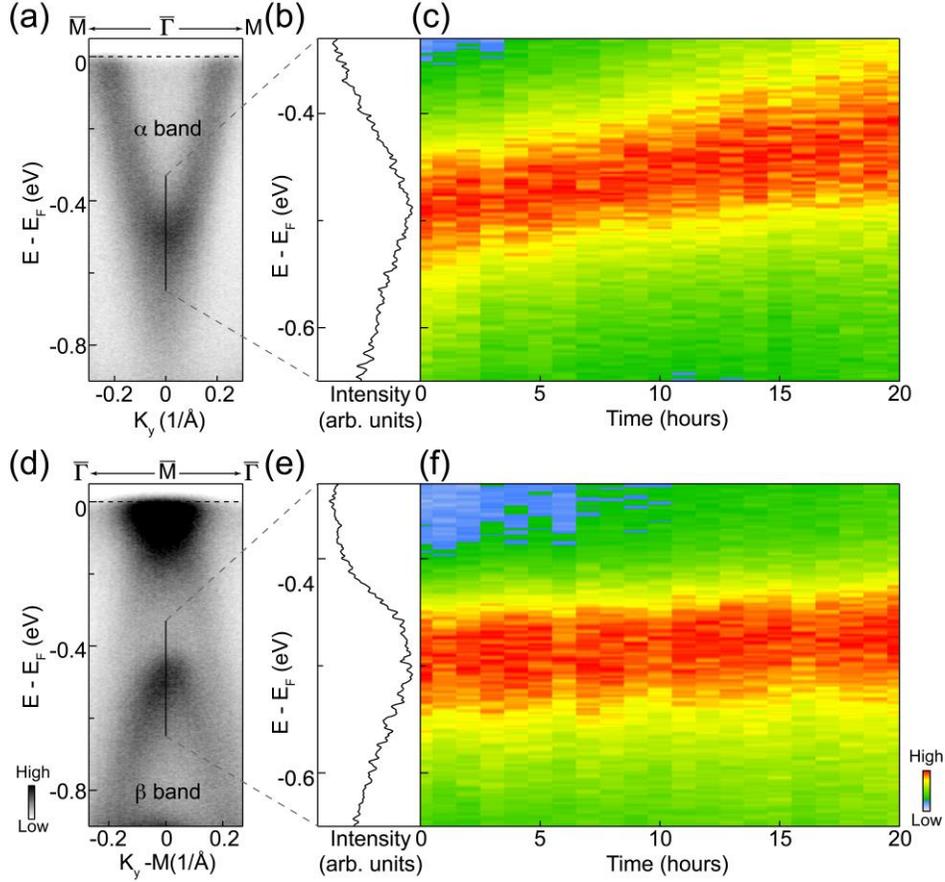

***Fig. 3 The evolution of band structure as a function of time.*** *(a) Photoelectron intensity plot of the $\alpha$ band around $\overline{\Gamma}$. (b) EDC at $\overline{\Gamma}$ with its energy range marked by the black line in (a). (c) Time evolution of the EDC in (b). The intensity of the EDC peak is shown by the color scale. (d-f) Same as (a-c), but for the $\beta$ band around $\overline{M}$.*

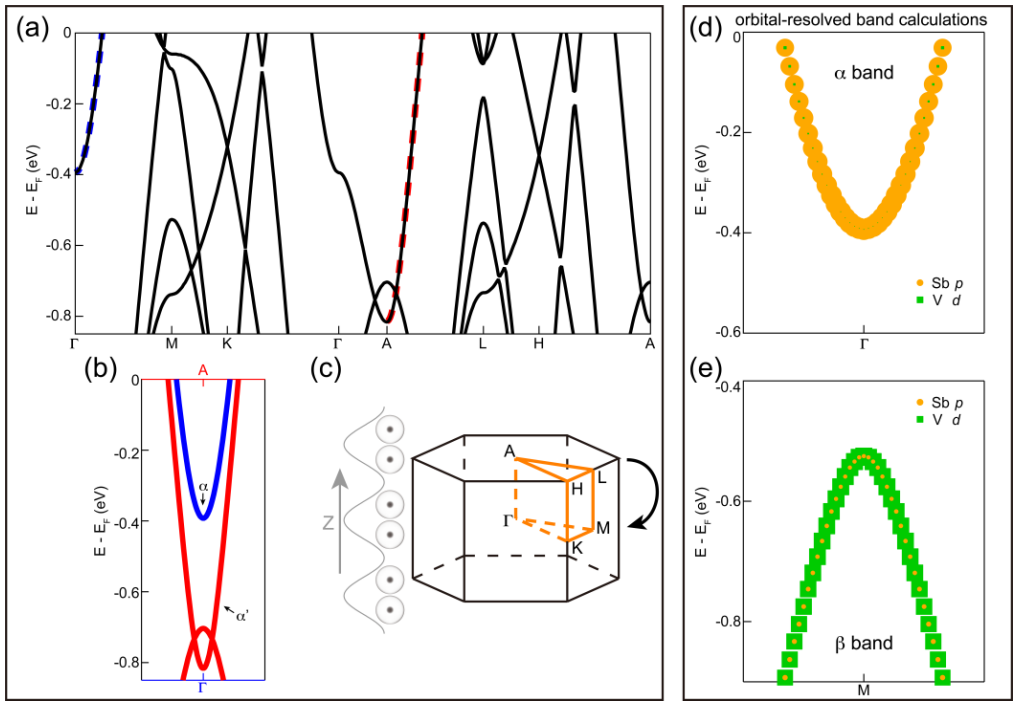

***Fig. 4 DFT calculations of the band structure.*** *(a) Calculated band structure along high symmetry directions in the 3D BZ. (b) The folded band structure, in which the bands along L-A-L (red) and M-Γ-M (blue) coexist along the same momentum cut. (c) Schematic of the BZ folding along the out-of-plane direction, induced by the 3D charge order. (d-e) Orbital-resolved calculations for α band (d) and β band (e). The size of the symbols represents the relative projection contribution of the corresponding orbitals.*

The origin of the band reconstruction as a function of time is necessarily associated with changes of sample surface; however, the observed band evolution is not due to a trivial aging effect. Sample aging typically broadens the measured spectrum but should not shift the energy of the band. The different evolution between $\alpha$ band and $\beta$ band observed in the experiment is also distinct from a simple band filling with carriers, which would give rise to a rigid band shift. To deepen the understanding, orbital-resolved DFT calculations have been carried out. As shown in Fig. 4d,e, the $\alpha$ band is dominated by Sb *p*-orbitals, whereas the $\beta$ band is mainly from V *d*-orbitals. Therefore, the time-dependent band evolution represents an orbital-selective energy shift of the electron energy band, which mainly involves the Sb *p*-orbitals. This is consistent with a recent proposal considering an orbital-selective hybridization of hole carriers from surface vacancies of Cs atoms [11], where the Sb $p_z$-orbitals pointing towards the Cs atoms are mainly involved. In the calculation, the $\alpha$ band shows an upward shift in energy with increasing Cs vacancies, whereas the $\beta$ band remains unchanged [11]. This result is well-aligned with our experimental observation, if we assume the fresh sample surface loses Cs atoms and forms Cs vacancies as a function of time. The formation of Cs vacancies on the cleaved sample surface has also been observed by STM measurements. Most importantly, this orbital-selectively reconstructed band structure could in turn enhance the superconductivity and suppress the charge order [11]. It indicates the feasibility of orbital-selective control of the band structure and physical properties in $CsV_3Sb_5$ via surface modification.

The last thing to discuss is the absence of the $\alpha'$ band when the sample is slowly cooled down to a low temperature below the charge ordering transition (Fig. 1e). One possibility is that the accumulated Cs vacancies break the long-range coherence of the 3D charge order on a few top-layers of the sample. STM results show that the pure long-range 2X2X2 charge order mainly appears on the surface region with periodic Cs atoms, whereas the Sb-terminated region is dominated by a 1X4 surface modulation [20,21]. If the $\alpha'$ band is induced by the 3D 2X2X2 charge order, then it may disappear on the sample surface without periodic Cs atoms.

Our work has identified two distinct band reconstructions in $CsV_3Sb_5$. The new band observed at low temperature is reproduced by considering a coupling with the long-range out-of-plane correlation of the 3D charge order. The orbital-selective band reconstruction observed as a function of time, on the other hand, is consistent with an orbital-selective hybridization of hole carriers from surface vacancies. It thus provides a unique window to manipulate the electronic structure and associated physical phenomena in this new class of materials. Future studies exploring surface manipulation with different dopants [43,44] are motivated by our result.

**Methods**

Single crystals of $CsV_3Sb_5$ were grown by the self-flux method as described elsewhere [1,2]. The samples were cleaved *in-situ* with a base pressure of 6 X $10^{-11}$torr. The photoemission measurements were carried out at our lab-based ARPES system with 21.2eV photons. The energy resolution was set at 5meV for the measurements. The Fermi level was referenced to that of a polycrystalline Au in electrical contact with the samples. DFT calculations were performed without spin-orbit-coupling in VASP *v*5.4.4 [45-47] using identical parameters to a recent study [2]. The calculated Fermi level was set to the DFT value and the orbital projections were determined using LOBSTER [48,49] for improved fidelity. While recent work has demonstrated a slight shift of the Fermi level relative to the DFT calculation, as well as energy gaps associated with spin-orbit coupling, these small adjustments do not affect the arguments discussed in the current study.

**Acknowledgements**

The work at university of science and technology of China (USTC) was supported by the Fundamental Research Funds for the Central Universities (No. WK3510000012) and USTC start-up fund. The work at PSI was supported by the Swiss National Science Foundation under Grant. No. 200021-188413, the Sino-Swiss Science and Technology Cooperation (Grant No. IZLCZ2-170075). The work at UC Santa Barbara was supported via the UC Santa Barbara NSF Quantum Foundry funded via the Q-AMASE-i program under award DMR-1906325. This research made use of the shared facilities of the NSF Materials Research Science and Engineering Center at UC Santa Barbara (DMR- 1720256). The UC Santa Barbara MRSEC is a member of the Materials Research Facilities Network (www.mrfn.org). S.M.L.T. acknowledges use of the shared computing facilities of the Center for Scientific Computing at UC Santa Barbara, supported by NSF CNS-1725797 and NSF DMR-1720256. B. R. O. acknowledges support from the California NanoSystems Institute through the Elings Fellowship program. S.M.L.T has been supported by the National Science Foundation Graduate Research Fellowship Program under Grant No. DGE-1650114.

**Author Contributions**

J.-F.H. designed the research. B.R.O. grew and characterized the crystals with support from S.D.W.. S.M.L.T. performed the theoretical calculations. Y.L. and L.H. performed the ARPES experiments with help from S.P., Z.W., J.S., Z.O., B.W., Y.M., M.G.. S.T.P. and Y.L. analyzed the data with inputs from Y.H. and M.S.. S.T.P. draw the figures. J.-F.H. and Y.L. wrote the paper with inputs from all authors.